\documentclass[twocolumn,prl,amsmath,amssymb,showpacs,superscriptaddress,floatfix]{revtex4}
\usepackage{graphicx}
\usepackage{bm}
\usepackage{lineno,hyperref}

\usepackage{epsf}
\usepackage{epstopdf}
\begin{document}
\title{Melting line and thermodynamic properties of a supeionic compound SrCl$_2$ by molecular dynamics simulation}

\author{Yu. D. Fomin}
\affiliation{ Institute for High Pressure Physics RAS, 108840
Kaluzhskoe shosse, 14, Troitsk, Moscow, Russia}

\date{\today}

\begin{abstract}
In the present paper we study the thermodynamic properties of superionic conductor $SrCl_2$ at high temperatures
by means of molecular dynamics method. Firstly, we calculate the melting line. Then we compute the equations
of state and the response functions (heat capacity, thermal expansion coefficient, etc) at the temperatures
up to the melting. We show that the response functions show maxima or minima at the temperatures well above the
temperature of transition into the conductive state, and therefore are not related to this transition.

\end{abstract}


\maketitle

Superionic conductors are crystalline ionic salts which demonstrate high electrical conductivity \cite{book1}. Their high electrical conductivity
is related to very different mobilities of two types of ions in the salt: while one type of the ions remains mostly
fixed at the lattice sites, another type demonstrates high diffusion. Depending on the substance the mobile type of particles
can be either cation or anion. Importantly, superionic compounds are widely used in modern energetics as solid electrolytes in the
batteries. Because of this deep understanding of their properties is of great importance for both fundamental research and technological
applications \cite{book2,book3}.

A wide class of superionic compounds is the one of salts with the structure of fluorite \cite{chadwick,disorder,fl1,fl2}. These substances have a
chemical formula $CA_2$ where $C$ means the cation and $A$ - the anion. In the fluorite structure the cations occupy
the cites of an Face Centered Cubic (FCC) lattice, while the anions are located in the tetrahedral holes. Numerous
salts with composition $CA_2$ demonstrate the fluorite structure, for instance, fluorite itself $CaF_2$, $BaCl_2$, $PbO_2$,
$SrCl_2$, etc. The mobile specie of the superionic compounds with fluorite structure is the anion.

At low temperatures the supeionic compounds do not demonstrate high conductivity. However, at certain temperature they experience
a phase transition into the conducting state. Typically phase transitions are related to some
thermodynamic peculiarities, such as the maximum of the heat capacity. The peak of heat capacity in the vicinity of
the transition into superionic state was reported in a number of experimental \cite{cp1,cp2} and theoretical \cite{cp3,cp4,cp5,cp6,gillan1}
works. Several theoretical models were proposed to describe the behavior of this peak. However, other thermodynamic response functions
were studied much less and therefore they still require a careful investigation.


In the present paper we study the thermodynamic properties of a superionic conductor $SrCl_2$ by means of molecular dynamics
method. We calculate the melting line and thermodynamic response functions (heat capacity, thermal expansion, etc).
We show that close to the melting line this substance demonstrates maxima or minima of several thermodynamic
properties. The temperatures of these peculiarities are rather far from the temperature of transition into conductive state.
Because of this we assume that these maxima and minima are not related to this transition, but are induced by some
other mechanisms, which means that the behavior of fast ions conductors is rather complex even far from the 
point of transition into the conductive state.

\bigskip


We simulate a system of $4116$ particles (i.e. 1372 atoms of Sr and 2744 atoms of Cl) in a cubic box with periodic
boundary condition by means of molecular dynamics methods. We use the Born-Mayer-Huggins potential
to describe the interaction between the species. It has the form

\begin{equation}\label{pot}
U_{\alpha \beta} (r) = \frac{Z_{\alpha}Z_{\beta} e^2}{r} +A_{\alpha \beta} e^{-r/\rho_{\alpha \beta}} -\frac{C_{\alpha \beta}}{r^6},
\end{equation}
where $\alpha$ and $\beta$ mark different species (Sr or Cl), $Z$ is the charge of the nuclei and $e$ is the elementary charge. Other
parameters of the potential are taken from Ref. \cite{gillan}. They are given in Table I. Importantly, Refs. \cite{gillan1,gillan} themselves
report a computational study of the behavior of $SrCl_2$. In particular, the authors also discuss the anomaly of the isochoric heat capacity. However,
they do not discuss the behavior of isobaric heat capacity and thermal expansion coefficient. They also do not calculate the melting line.

All simulations of the present work are performed using lammps simulation package \cite{lammps}.

\begin{table}
\begin{tabular}{|c|c|}
  \hline
  $A_{Sr-Sr}$ (eV) & 0.0 \\
  \hline
  $A_{Sr-Cl}$ (eV) & 774.14 \\
  \hline
  $A_{Cl-Cl}$ (eV) & 1227.2 \\
  \hline
  $\rho_{Sr-Sr}$ & - \\
  \hline
  $\rho_{Sr-Cl}$ ($\AA$) & 0.3894 \\
  \hline
  $\rho_{Cl-Cl}$ ($\AA$) & 0.3214 \\
  \hline
  $C_{Sr-Sr}$ (($ eV\AA^6$)) & 0.0 \\
  \hline
  $C_{Sr-Cl}$ (($ eV\AA^6$)) & 0.0 \\
  \hline
  $C_{Cl-Cl}$ (($ eV\AA^6$)) & 1.69 \\
  \hline
  \end{tabular}

\caption{Parameters of potential of Eq. \ref{pot} used in the present study. Note, that since $A_{Sr-Sr}$ is zero, the parameter $\rho_{Sr-Sr}$ is not required.}
\end{table}



\begin{figure}
\includegraphics[width=6cm,height=6cm]{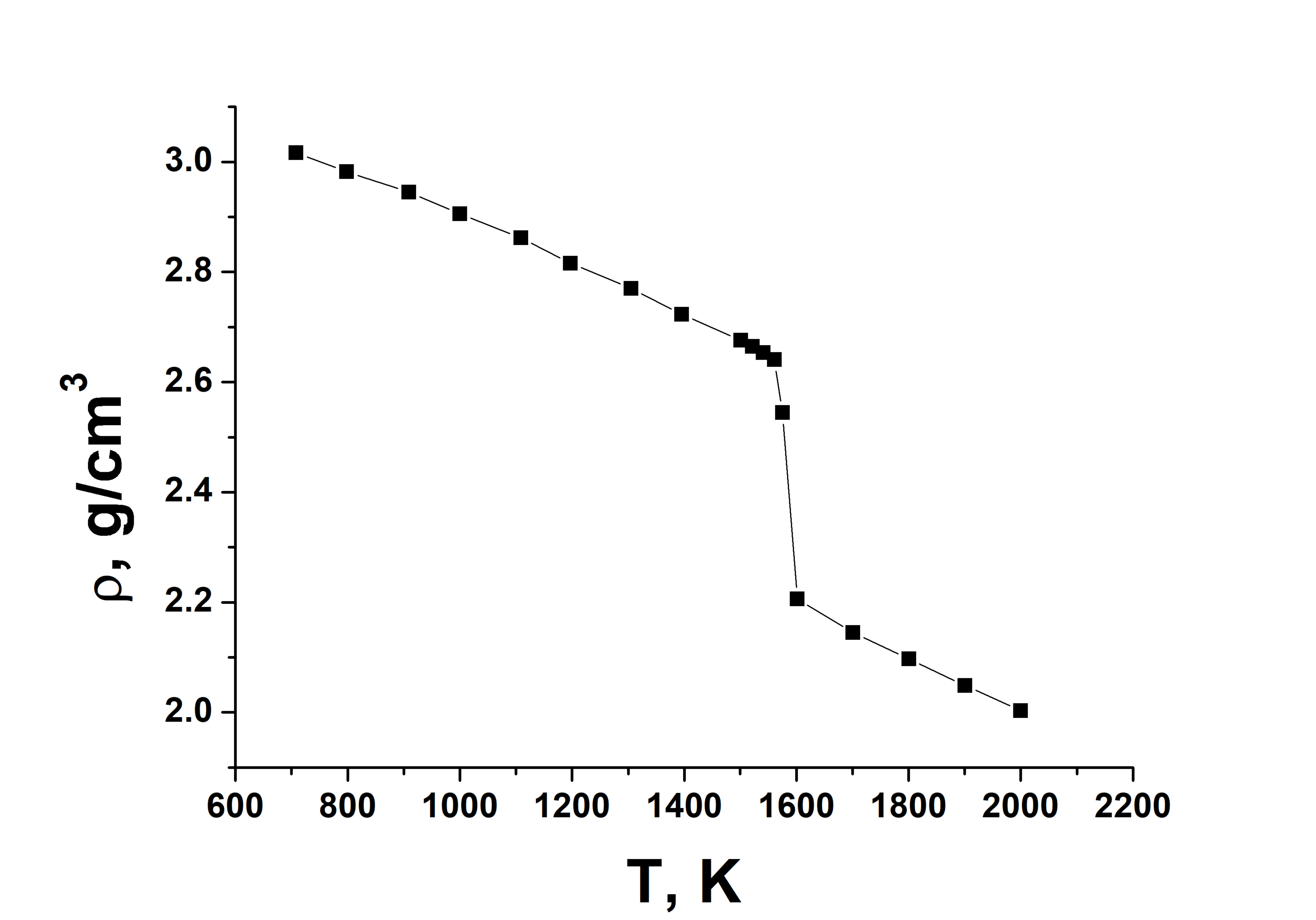}%

\caption{\label{p1} Equation of state of $SrCl_2$ at ambient pressure and different temperatures. The jump
of the density signalizes that melting of the system takes place.}
\end{figure}

We start our discussion from simulation of $SrCl_2$ at ambient pressure and different temperatures. The system is simulated
at fixed pressure (P=1 bar) and different temperatures for 10 ns with the time step 1 fs. In these simulations we
evaluate the average density of the system and monitor the structure.
The dependence of the density on the temperature is shown in Fig. \ref{p1}. One can see that the
density experiences a jump at $T=1560$ where it is $\rho=2.64$ $g/cm^3$. After the jump at $T=1600$ K
the density is $\rho=2.2$ $g/cm^3$. From investigation of the structural properties
we observe that at low temperatures the system is solid, while at the high ones it is liquid.
Although this method does not give accurate estimation of the melting point,
it allows us to estimate the lowerst density of solid $SrCl_2$ to be studied.

\begin{figure}
\includegraphics[width=6cm,height=6cm]{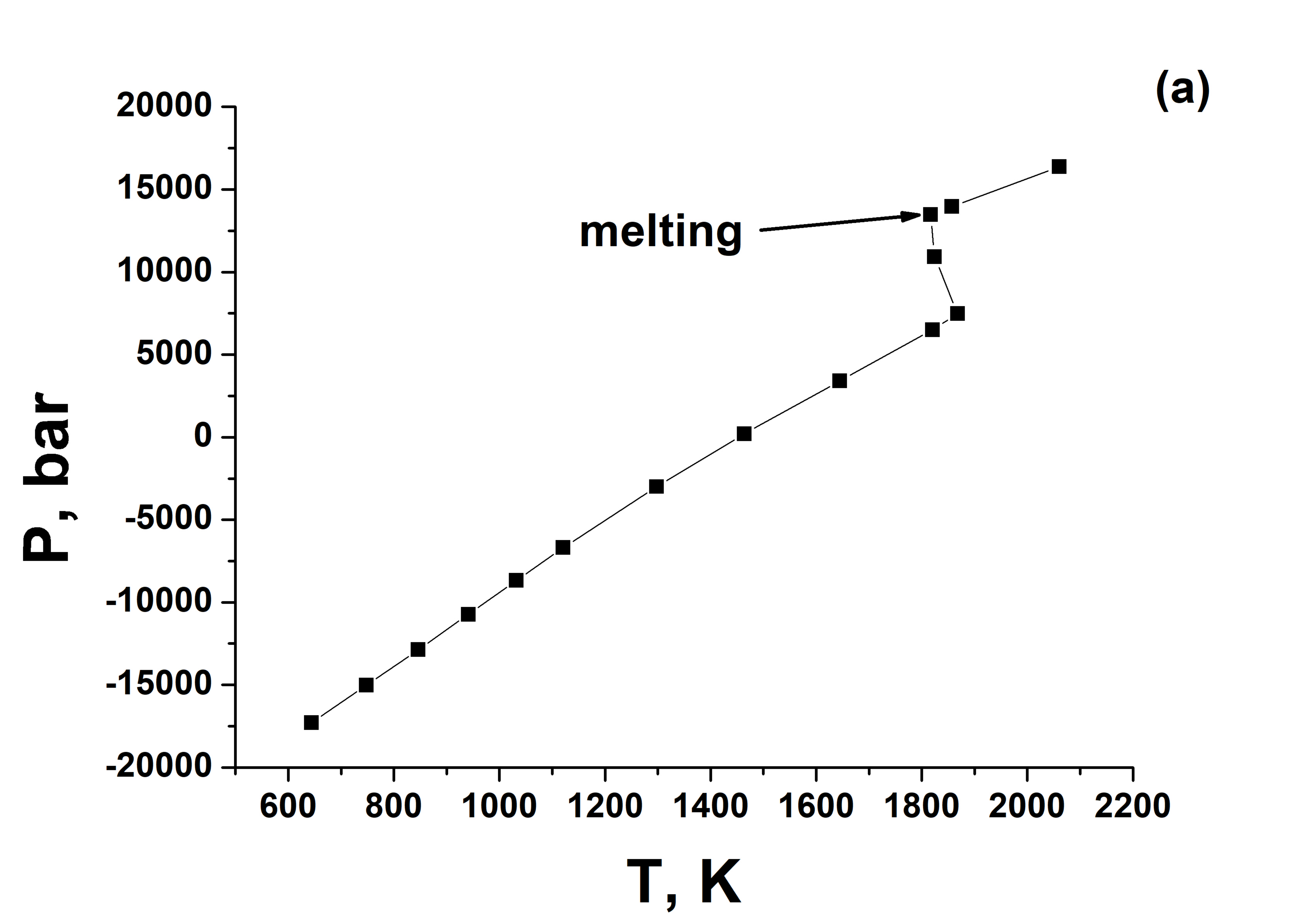}%

\includegraphics[width=6cm,height=6cm]{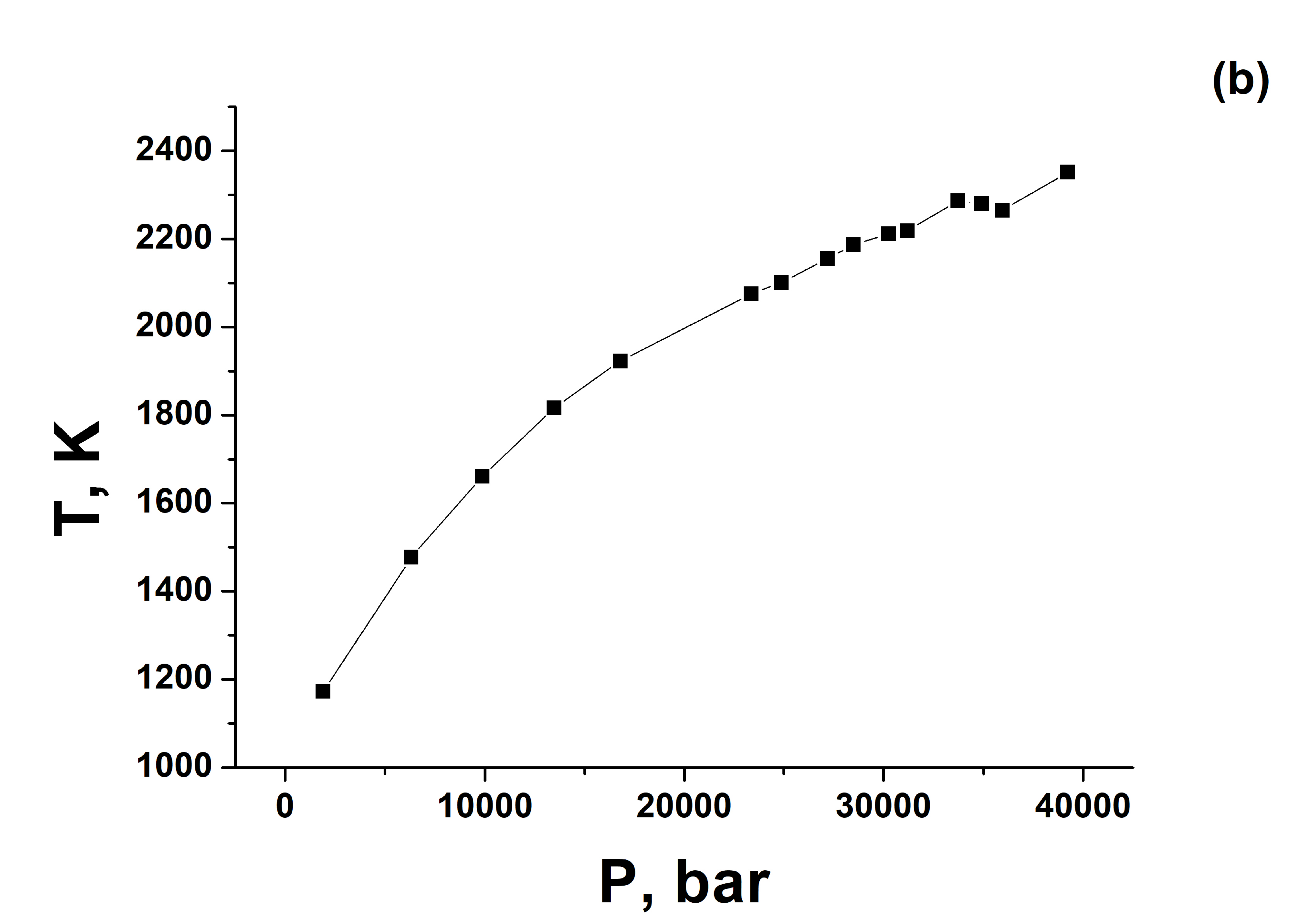}%

\caption{\label{melting-line} (a) Example of calculations of the melting point by Z-method. $\rho=2.7$ $g/cm^3$.
(b) Melting line of $SrCl_2$ obtained by Z-method calculations.}
\end{figure}

Having estimated the melting density at ambient pressure, we calculate the melting line on a more precise ground.
For doing this we employ a so called Z-method \cite{zmethod}. The density is varied from $\rho_{min}=2.6$ $g/cm^3$
up to $\rho_{max}=3.05$ $g/cm^3$. The system is simulated for 10 ns with the time step 1 fs
in microcanonical ensemble (constant number of particles N, energy E and volume V). According to the Z-method,
the dependence of pressure on the temperature has a Z-like shape at the melting point.
Fig. \ref{melting-line} (a) shows an example of Z-method calculations for $\rho=2.7$ $g/cm^3$.
One can see that the melting temperature is $T_m=1817$ K and the melting pressure is $13483$ bar. The melting line
is given in Fig. \ref{melting-line} (b).

\begin{figure}
\includegraphics[width=6cm,height=6cm]{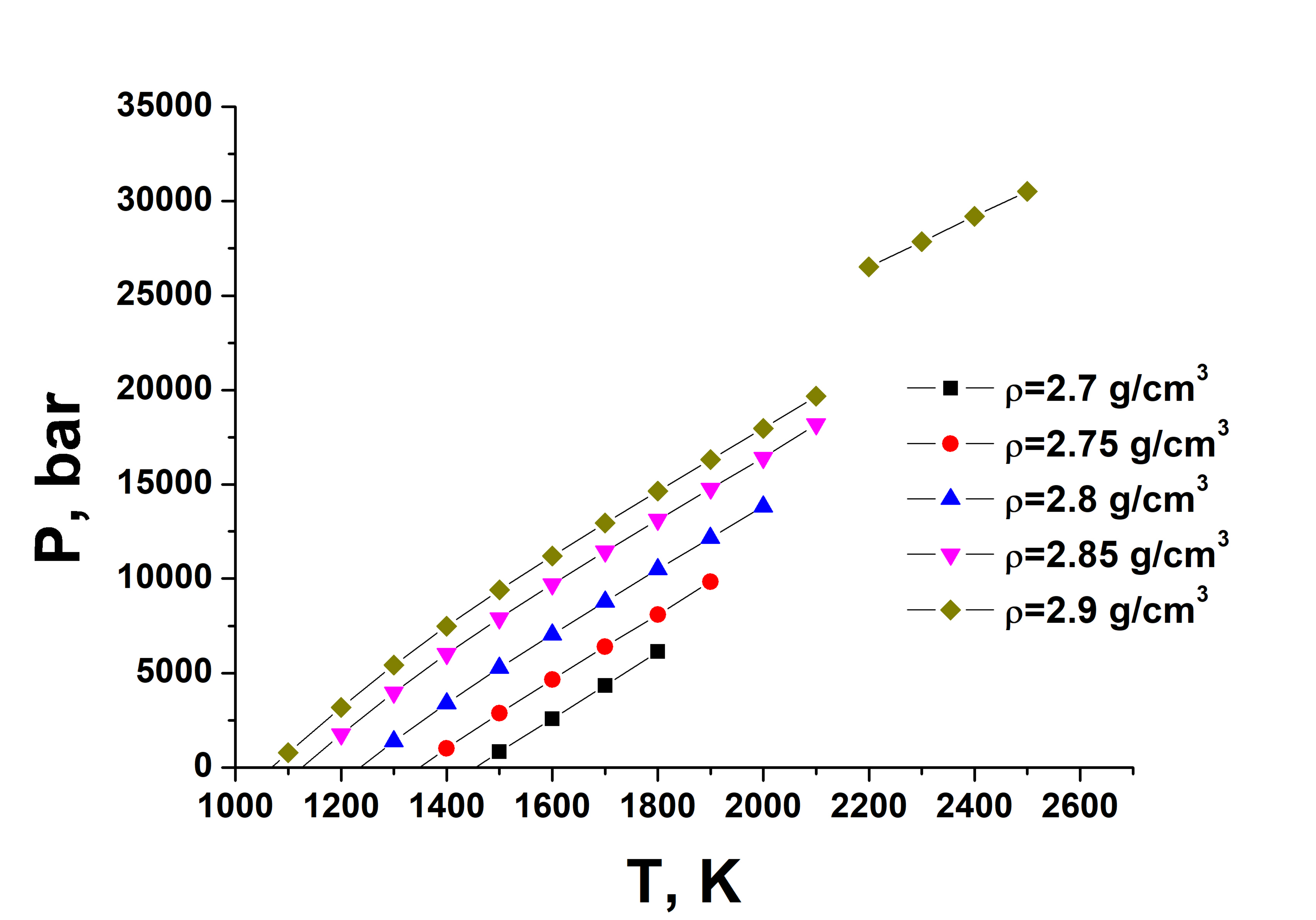}%

\caption{\label{isov} Equation of state of $SrCl_2$ along several isochores. Jump at the isochore
$\rho=2.9$ $g/cm^3$ signalizes that melting of the system takes place. The points at the temperature
above the jump correspond to the liquid state while below to solid. The data for liquid for other isochores
are removed from the plot to make it clearer.
}
\end{figure}

We calculate equation of state of $SrCl_2$ along several isochores from $T_{min}=1000$ K up to $T_{max}=2600$ K. Examples of
these equations of state are given in Fig. \ref{isov} (a). At some points the pressure has negative values. These points are
discarded from the plot. A jump of equation of state (shown only for $\rho=2.9$ $g/cm^3$) means crossing of the melting
line. The points corresponding to liquid are also discarded since the focus of the present study is
related to the crystalline state only. These equations of state do not demonstrate any unusual features: pressure
monotonously increases with temperature.


\begin{figure}
\includegraphics[width=6cm,height=6cm]{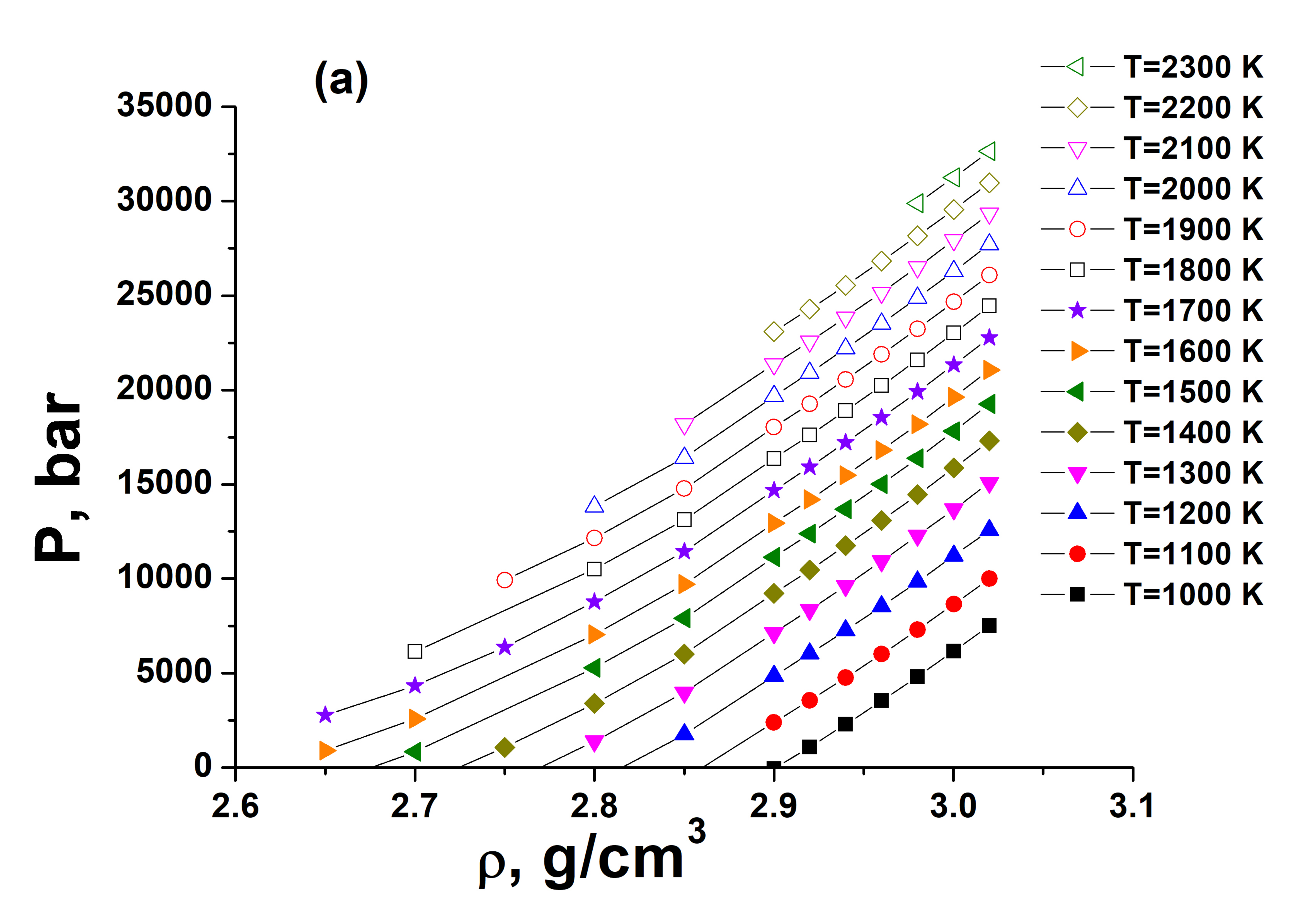}%

\includegraphics[width=6cm,height=6cm]{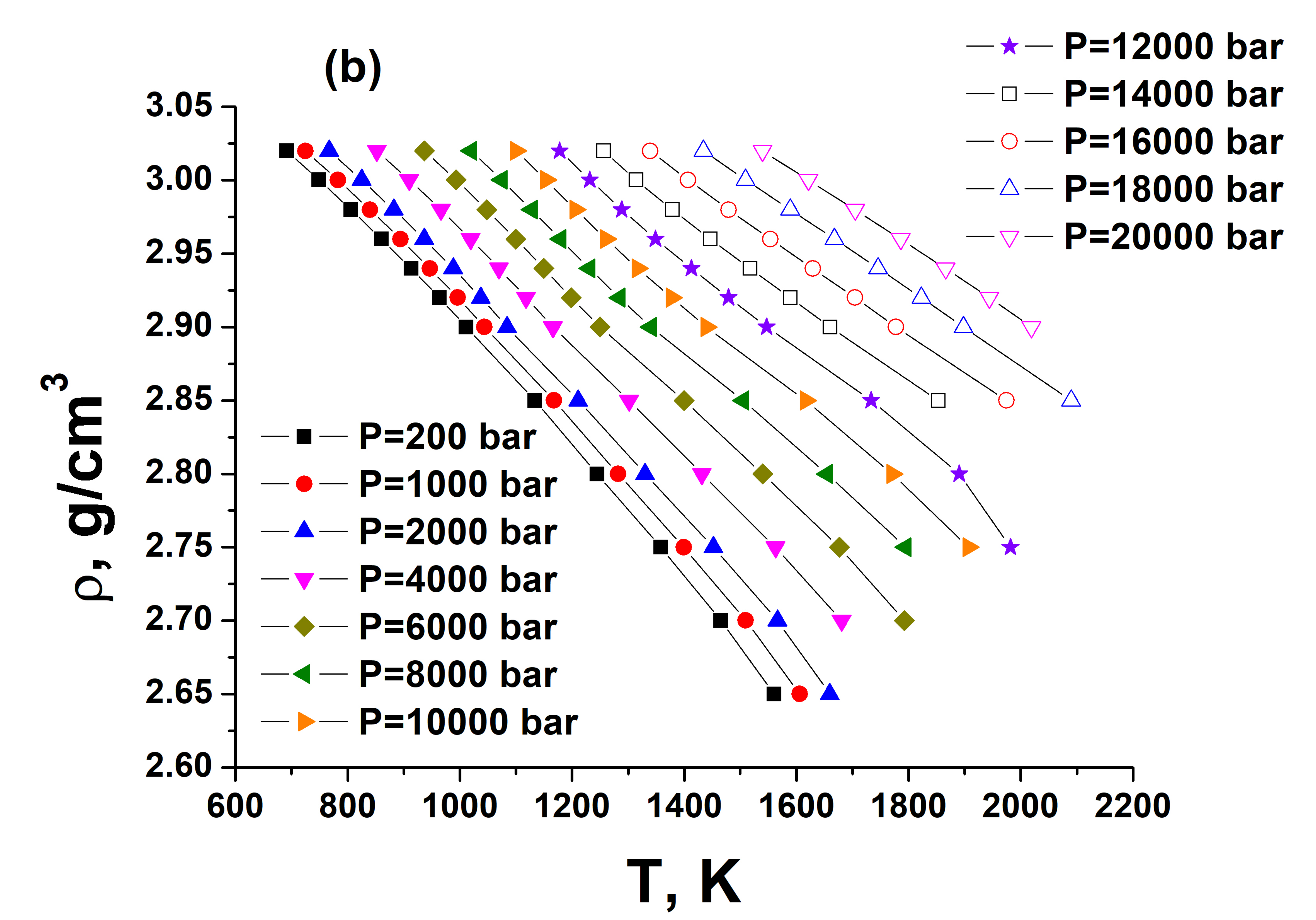}%

\caption{\label{isopt} (a) Equation of state of $SrCl_2$ along several isotherms. (b) Equation of state along
several isobars.}
\end{figure}

In our study we simulate almost 200 data points in ($\rho$,T) plane along several isochores. It allowed us
to find the properties of the system along other thermodynamic pathes like isotherms and
isobars by interpolation of our data. Simple linear interpolation was enough to obtain the data along
isotherms and isobars. Numerical differentiation of the pressure, internal energy or enthalpy was
used to calculate the thermodynamic response functions. Fig. \ref{isopt} shows the equation of state of
solid $SrCl_2$ along several isotherms and isobars.

\begin{figure}
\includegraphics[width=6cm,height=6cm]{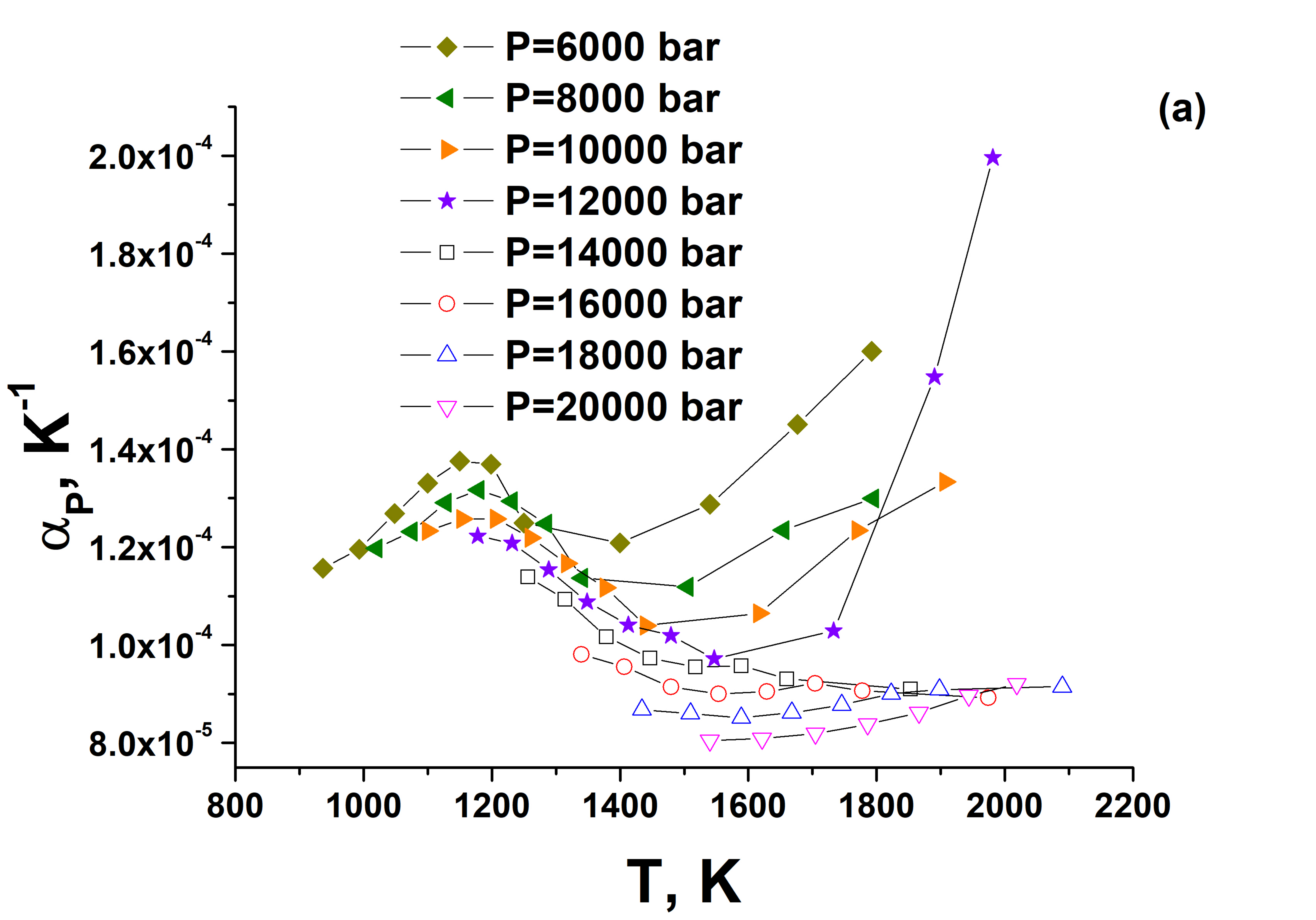}%

\includegraphics[width=6cm,height=6cm]{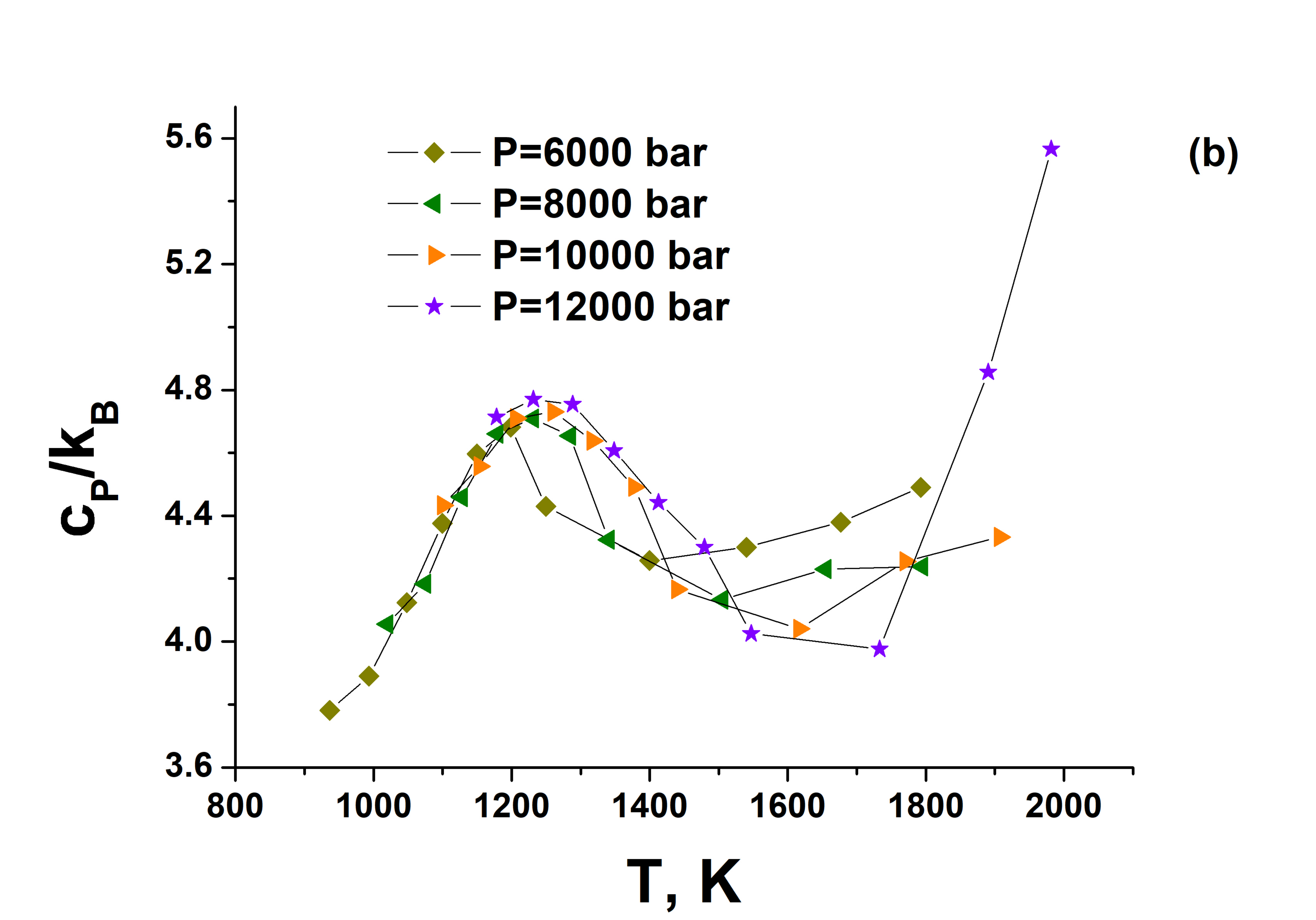}%

\includegraphics[width=6cm,height=6cm]{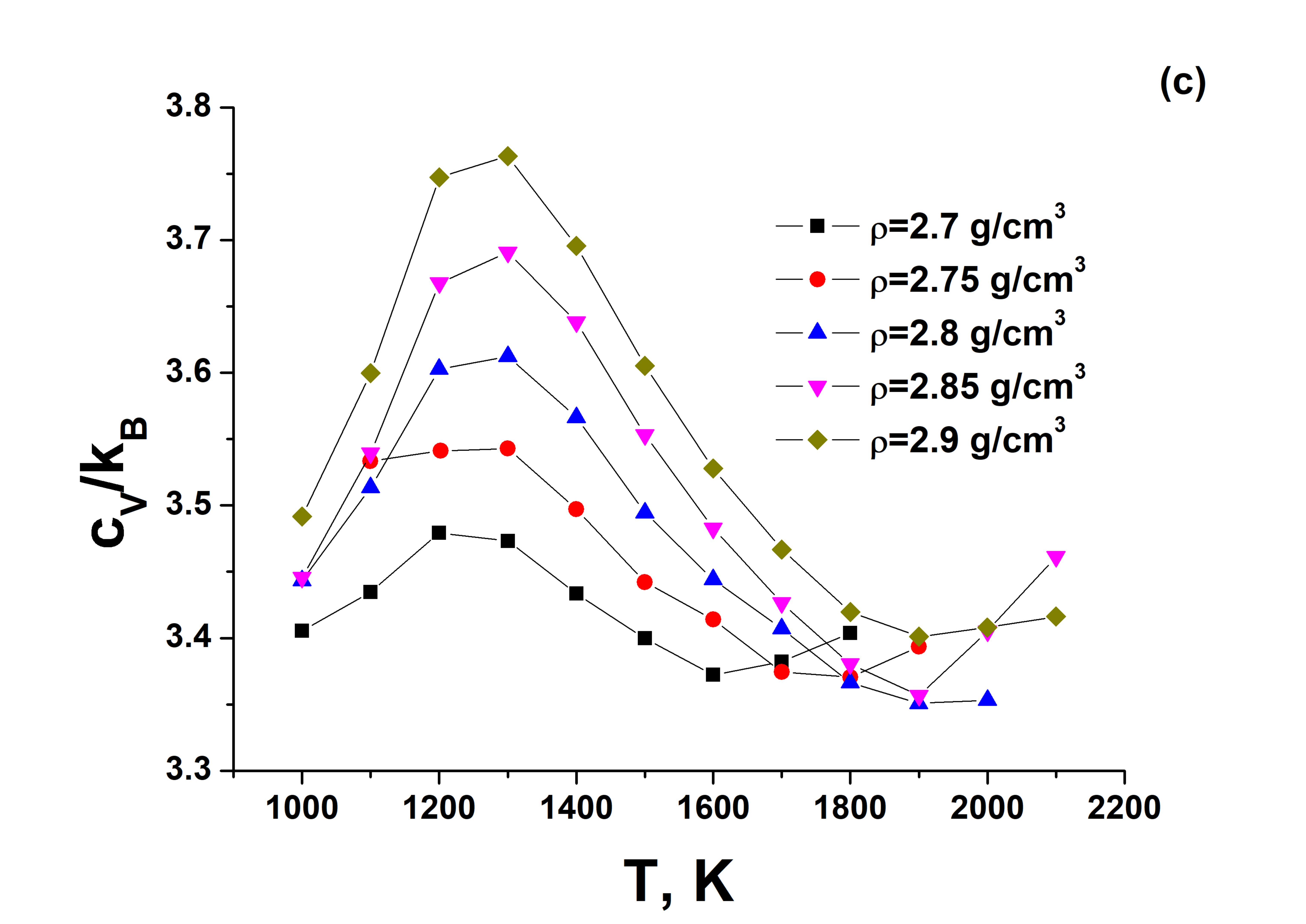}%

\caption{\label{response} (a) Thermal expansion coefficient of $SrCl_2$ along several isobars. (b) Isobaric heat capacity along
several isobars. (c) Isochoric heat capacity along several isotherms.}
\end{figure}

Although the equation of state does not demonstrate any peculiarities neither along isobars, not isotherms and isochores, the thermodynamic
response functions do demonstrate some maxima and minima. We calculate the thermal expansion coefficient
$\alpha_P= \frac{1}{V} \left( \frac{\partial V}{\partial T} \right)_P$, the isochoric heat capacity $c_V=\left( \frac{\partial U}{\partial T} \right)_V$,
where $U$ is the internal energy of the system, and the isobaric heat capacity $c_P=\left( \frac{\partial H}{\partial T} \right)_V$, where
$H=U+PV$ is the enthalphy of the system. The results of these calculations are given in Fig. \ref{response} (a)-(c). One can see that
all three functions demonstrate maxima and minima.

The location of maxima of response functions is widely discussed in the vicinity of liquid-gas or liquid-liquid critical point
\cite{w1,w2,w3,w4,w5,w6,w7,w8,w9}. In these cases the locations of maxima in density-temperature (or, equivalently, pressure-temperature) plane
should tend to the critical point as the temperature tends to the critical one. However, the curves of maxima of different quantities
rapidly diverge as the temperature goes away from the critical point.

\begin{figure}
\includegraphics[width=6cm,height=6cm]{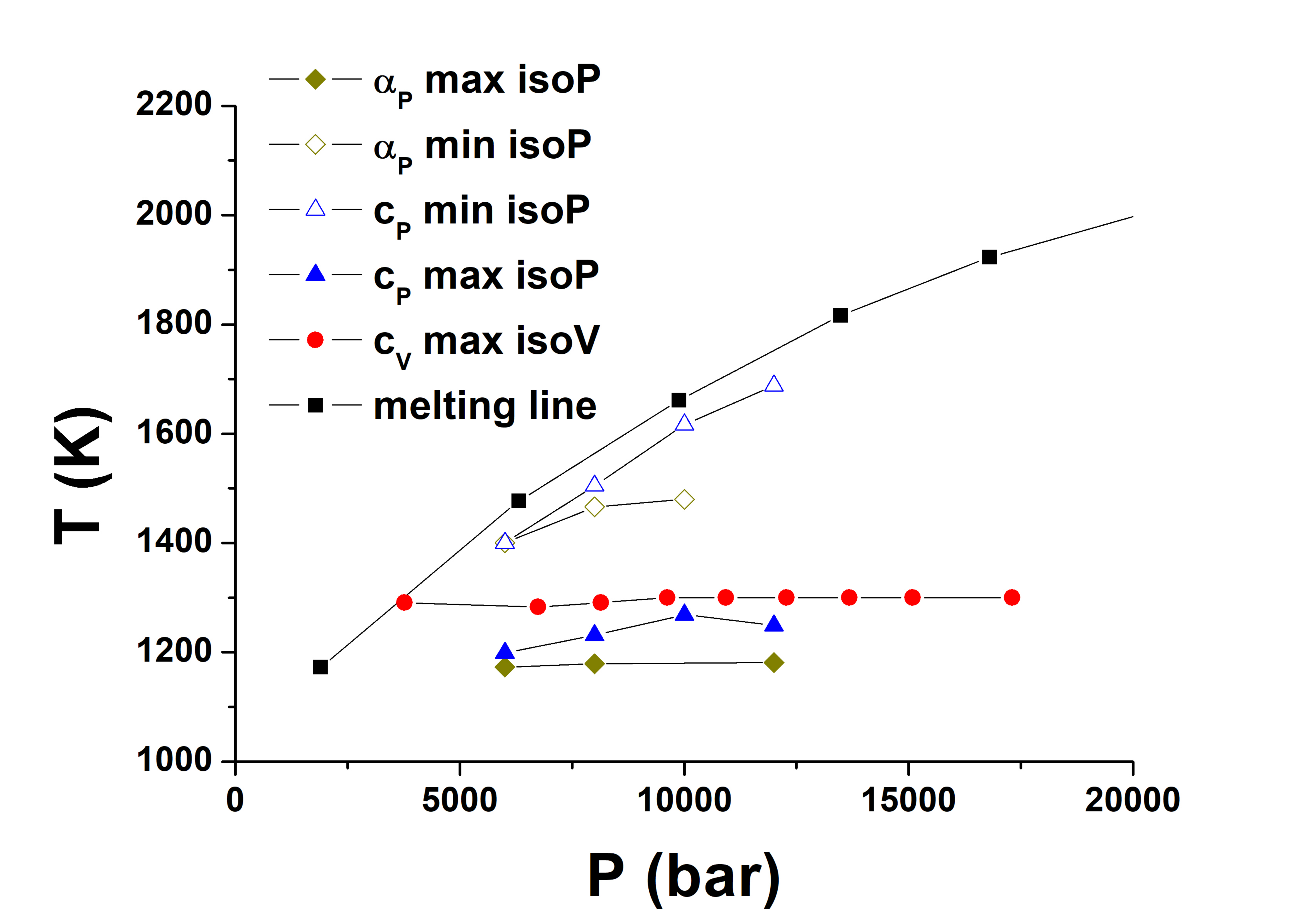}%

\caption{\label{max-pd} Maxima and minima of the response functions of $SrCl_2$ placed in the phase diagram of the system.}
\end{figure}

In the case of $SrCl_2$ the situation is different and one does not expect coincidence of the lines of maxima of different functions.
However, the location of the maxima and minima of the response functions is of a great interest for characterisation of the
thermodynamic properties of the system. Fig. \ref{max-pd} shows the points of maxima and minima of $\alpha_P$, $c_P$ and $c_V$
on the phase diagram of the system. One can see that the lines of maxima of $c_V$ along isochores and minima of $\alpha_P$
along isobars are almost isotherms with $T_{c_V-max}=1300$ K and $T_{\alpha_P-min}=1180$ K. The line of minima of $c_P$
along isobars is located close to the melting line and goes almost parallel to it.

\begin{figure}
\includegraphics[width=6cm,height=6cm]{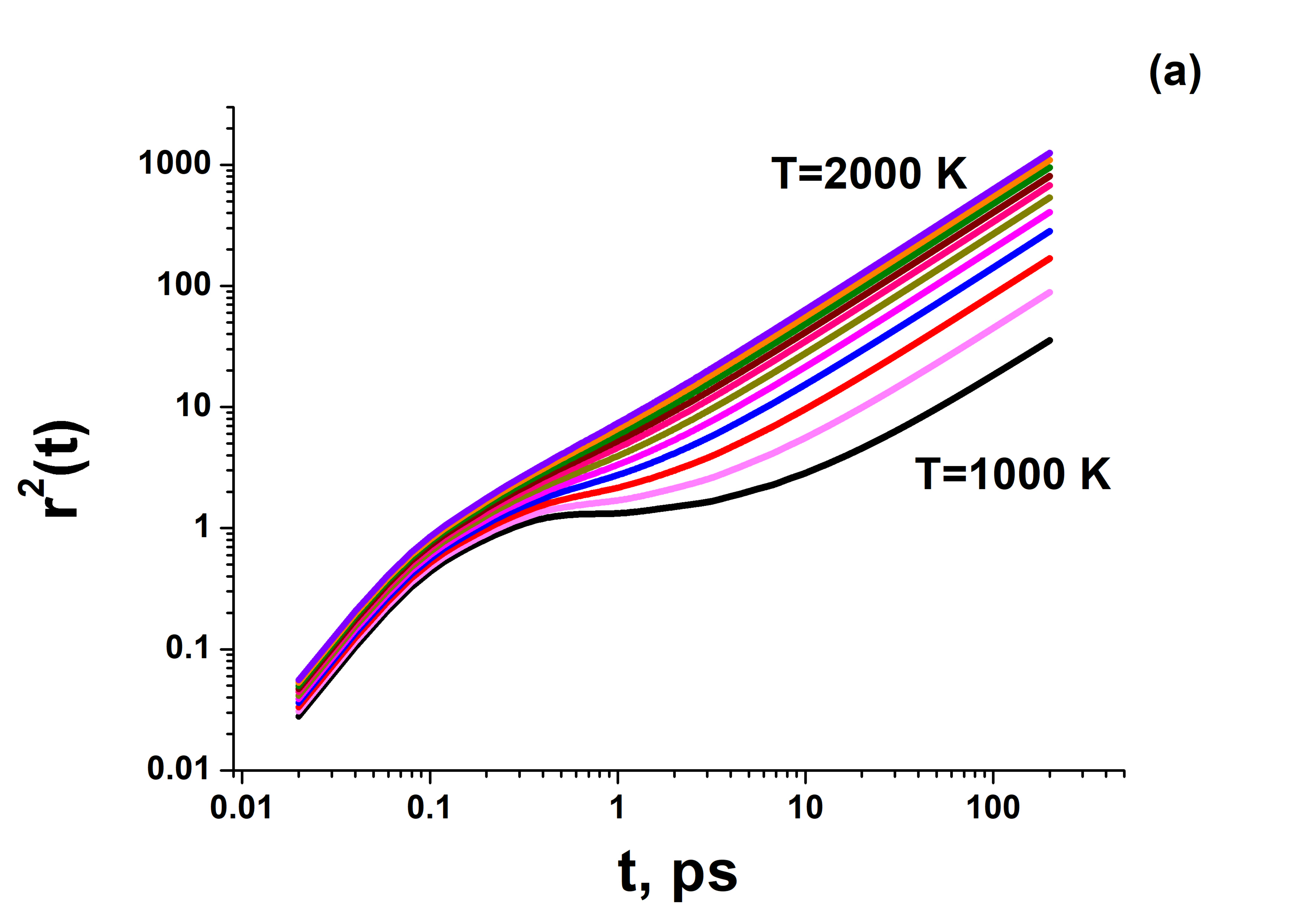}%

\includegraphics[width=6cm,height=6cm]{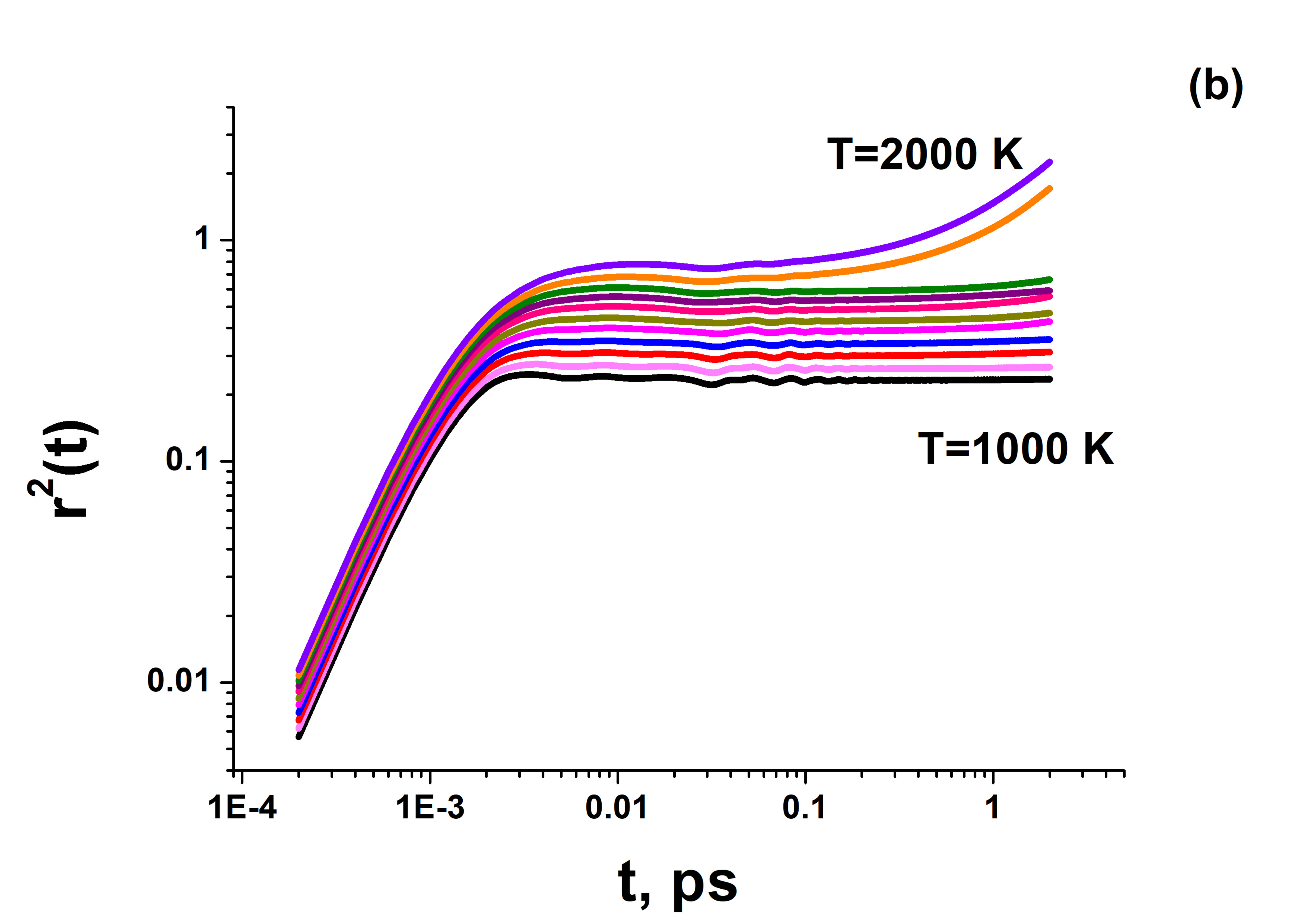}%

\caption{\label{diff} Mean square displacement of (a) clorine and (b) strontium at $\rho=2.8$ $g/cm^3$ and
different temperatures. In both cases the temperatures from $T=1000$ K (the lowest curve) to $2000$ K (the highest curve)
with the step $\Delta T=100$ K are shown}
\end{figure}

\begin{figure}
\includegraphics[width=6cm,height=6cm]{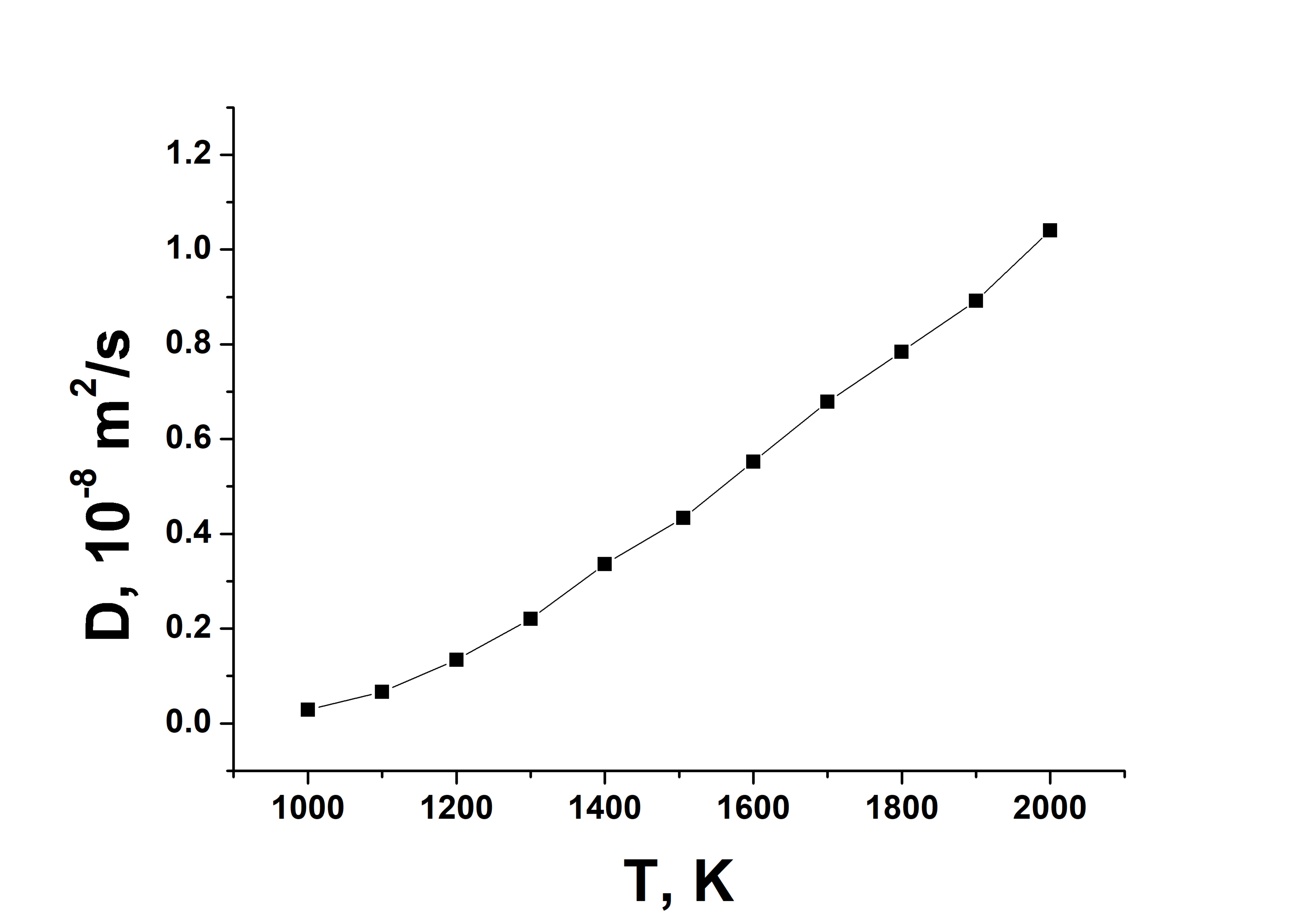}%

\caption{\label{diff1} Diffusion coefficient of chlorine at the isochore $\rho=2.8$ $g/cm^3$.}
\end{figure}

Finally we evaluate the diffusion coefficients of the species at different temperatures. Fig. \ref{diff} (a) and (b) show
the mean square displacement of chlorine and strontium at the density $\rho=2.8$ $g/cm^3$ and the temperatures from
1000 K to 2000 K, i.e. up to the melting point. One can see that even at $T=1000$ K the chlorine
has finite diffusion coefficient, the diffusion coefficient of strontium at this temperature is apparently zero. Therefore,
the system is in conducting state at this temperature. This is consistent with Ref. \cite{gillan}, where the transition
from insulating state into the conducting one is estimated to be about 1000 K. The mean square displacement of strontium
remains shoulder-like up to the temperature $T=1900$ K where a slight deviation at long times appears. This temperature
is close to the melting point and therefore moderate diffusion of $Sr$ due to the formation of defects can appear in the system.
Fig. \ref{diff1} shows the diffusion coefficient of chlorine at $\rho=2.8$ $g/cm^3$ and different temperatures. The diffusion of
Sr is nearly zero at all these temperatures.

One can see that the transition from insulating to conducting state appears at the temperatures much lower then the ones where we observe maxima
and minima of the response functions. At the same time some of them are also far from the melting line. Because of this it is not
evident whether one can relate the maxima and minima of the response functions to one of these transitions. The exact nature of
these maxima and minima is still under question and requires further investigation.

\bigskip

In conclusion, we performed a molecular dynamic study of thermodynamic properties of the melting line and thermodynamic properties of
superionic conductor $SrCl_2$. We calculated the melting line and the location of maxima and minima of thermal expansion
coefficient, isobaric and isochoric heat capacities. We find that the peculiarities of the thermodynamic functions are
located far from both phase transitions in the density-temperature (or pressure-temperature) plane. Because of this they
cannot be directly attributed to these transitions. The reasons for these peculiarities to appear require further investigation.

\bigskip

This work was carried out using computing resources of the federal
collective usage center "Complex for simulation and data
processing for mega-science facilities" at NRC "Kurchatov
Institute", http://ckp.nrcki.ru, and supercomputers at Joint
Supercomputer Center of the Russian Academy of Sciences (JSCC
RAS). The work was supported by the Russian Foundation of Basic Research (Grant No 18-02-00981).

\end{document}